\documentclass[a4paper,10pt,notitlepage]{article}
\usepackage{amsmath,amssymb,amsfonts,dcolumn,color,graphicx,graphics,latexsym,placeins,epsfig}
\usepackage{graphicx}  % standard LaTeX graphics tool,
\usepackage{amsmath}   % For getting proper math eqns
\usepackage{amssymb}   % Used for getting various symbols eg. gtrsim, lesssim etc.
\usepackage{bm} % For bold math (esp of lower greek letters)
\usepackage{dcolumn}% Align table columns on decimal point
\usepackage{color}
\usepackage{footmisc}
\usepackage{mathrsfs}
\usepackage{amsfonts}
\usepackage{varioref}
\usepackage{slashed}
\usepackage{footnote}
\usepackage[belowskip=-17pt,aboveskip=15pt]{caption}
\usepackage{amsmath}
\usepackage{makeidx}
\usepackage{amssymb}
\usepackage{graphicx}
\usepackage{bmpsize}
\usepackage[margin=1.45in]{geometry}
\usepackage{graphicx}
\usepackage{dcolumn}
\usepackage{amsmath,amssymb,mathtools}
\usepackage{epstopdf}
\usepackage{verbatim}
\usepackage[colorlinks=true, citecolor=blue, urlcolor = blue, linkcolor= blue, 
bookmarks=true]{hyperref}
\def\e{\epsilon}
\def\p{\partial}

\def\be{\begin{equation}}
\def\ee{\end{equation}}

\labelformat{section}{Section #1} 
\labelformat{subsection}{Section #1} 
\labelformat{subsubsection}{Section #1}
\labelformat{subsubsubsection}{Section #1}
\labelformat{equation}{Eq.~(#1)} 
\labelformat{figure}{Fig.~#1} 
\labelformat{subfigure}{Fig.~\thefigure#1} 
\labelformat{table}{Tab.~#1} 
\labelformat{appendix}{Appendix #1}
\begin{document}
%%%%%%%%%%%%%%%%%%%%%%%%%%%%%%%%%%%%%%%%%%%%%%%%%%%%%%%%%%%%%%%%%%%%

%%%%%%%%%%%%%%%%%%%%%% TITLE PAGE %%%%%%%%%%%%%%%%%%%%%%%%%%%%%%%%%%%
\thispagestyle{empty}
\setcounter{page}{0}
\begin{flushright}\footnotesize
%\texttt{DESY }\\
%\texttt{HU-Mathematik-}\\
%\texttt{HU-EP-}\\
\vspace{0.5cm}
\end{flushright}
\setcounter{footnote}{0}

\begin{center}
{\LARGE{
\textbf{Decoherence and entropy generation at one loop in the inflationary de Sitter spacetime for Yukawa interaction} 
}}

\medskip

{\large
$^{1}$Sourav Bhattacharya\footnote{sbhatta.physics@jadavpuruniversity.in}\, and\, $^2$Nitin Joshi\footnote{2018phz0014@iitrpr.ac.in}\\
\vskip .4cm
\small{$^1$Relativity and Cosmology Research Centre, Department of Physics, Jadavpur University, Kolkata 700 032, India}\\
\small{$^2$Department of Physics, Indian Institute of Technology Ropar, Rupnagar, Punjab 140 001, India}\\}

%\textbf{Abstract}\\[2mm]
\end{center}
\begin{abstract}
\noindent
The decoherence mechanism is  believed to be possibly connected to the quantum to classical transition of the primordial cosmological perturbations in the early universe. In this paper, we extend our previous analysis on decoherence in a fermion and scalar quantum field theory coupled via the Yukawa interaction in the Minkowski spacetime, to the inflationary de Sitter background. We treat the scalar field as the system and the fermions as the environment, and both the fields are taken to be massless. We utilise a non-equilibrium effective field theory formalism,  suitable for open quantum systems such as this.  We assume that an observer measures only the Gaussian 2-point correlator for the scalar field, as the simplest realistic scenario.
In order to compute the von Neumann entropy generated at late times as a measure of the  decoherence, we construct the one loop renormalised Kadanoff-Baym equation, which is the equation of motion satisfied by the 2-point correlators in the closed time path Schwinger-Keldysh formalism. These equations account to the self energy corrections. Using this, we next construct the one loop corrected statistical propagator for the scalar, which is related to its phase space area, to compute the von Neumann entropy.
We also compute the variation of the von Neumann entropy with respect to  relevant parameters. We note the qualitative similarity between our findings and the scenario where both the system and the environment are scalars. Our result is also qualitatively similar to an earlier one found by using the influence functional technique for a massive Yukawa theory. 
\end{abstract}


\newpage
\setcounter{page}{1}

%%%%%%%%%%%%%%%%%%%%%%%%%%%%%%%%%%%%%%%%%%%%%%%%%%%%%%%%%%%%%%%%%%%%
%%%%%%%%%%%%%%%%%%%%%%%%%%%%%%%%%%%%%%%%%%%%%%%%%%%%%%%%%%%%%%%%%%%%

\tableofcontents
\addtolength{\baselineskip}{5pt}

\section{Introduction}

The proposition  of the primordial cosmic  inflationary epoch as described in various instances \cite{aguth, aguth1, Weinberg}, has been highly successful in explaining the observed high degree of  spatial homogeneity of the Universe at very large scales.  Inflation  can also successfully explain the formation of structures we observe today,  starting from primordial quantum fluctuations. These predictions have been supported by observations of cosmic microwave background anisotropies, as seen in the  data from the WMAP \cite{Spergel} and Plank \cite{ade, Aghanim:2015xee, Aghanim:2018eyx, Akrami:2019bkn} missions. The inflationary phase is usually characterised by the dynamics of a scalar field called inflaton, which slowly rolls down a potential and leads to a nearly inflationary de Sitter phase, i.e. exponential expansion. For an excellent pedagogical discussion on various aspects of inflation and original references, we refer our reader to e.g.~\cite{Weinberg}.

It is well known that massless yet conformally non-invariant quantum fields, such as gravitons and a massless minimally coupled scalar  demonstrate strong infrared temporal growth at late cosmological times in the de Sitter spacetime, known as the  secular effect. These effects are  characterised by the logarithm of the scale factor and are caused by the   created long lived virtual particles living inside loops in the deep infrared, or super-Hubble limit, indicating a breakdown of the perturbation theory at late times~\cite{nitin, Bhattacharya:2022aqi, Friedrich:2019hev, Onemli:2002hr, Tsamis:2005hd, Cabrer:2007xm, Glavan:2021adm, Akhmedov:2013xka, Akhmedov:2014doa, Akhmedov:2015xwa, Akhmedov:2019cfd, Kaplanek:2020iay, Hu:2018nxy}.  Quantum fields that exhibit such secular effects are often regarded as the  {\it spectator} field. If a spectator field interacts with a normal matter field, due to strong quantum effects, the latter might also inherit such effects. As was pointed out in \cite{Friedrich:2019hev}, this feature changes the correlator and hence the degree of decoherence/classicalisation non-trivially.  Putting these in together, it seems an interesting task  to study the decoherence or quantum-to-classical transition in the inflationary background, in the presence of such spectator fields. 

Decoherence is a process in which the states of an open quantum system become entangled (via some interaction) with its surrounding which leads to the loss of coherence and correlation for the system. It has been gaining interest in the research community in the context of interacting quantum field theories~\cite{Calzetta Hu, Calzetta Hu1} (also references therein). We refer our reader to~\cite{ gravity, gravity1, acceleration, darkmatter, cosmology, cosmology1, DC, DC1, DC2, DC3, Janssen:2007ht, Markkanen:2016vrp, Hu:1992xp, Hu:1990cr} and references therein for vast study of quantum decoherence in the context of, for example, weak stochastic gravitational perturbation, accelerated time-delay source for an inertial observer, due to  the gravitational interaction of the dark matter with its environment and the classicalisation of the primordial inflationary quantum field theoretic perturbations. We further refer our reader to \cite{Brandenberger:1992sr, Polarski:1995jg, Lesgourgues:1996jc, Brandenberger:1990bx, Kiefer:1998qe, Prokopec:2006fc, Zurek:2003zz, Weenink:2011dd} for relevant discussions. Quantum field theory in a time dependent background such as the de Sitter is essentially a non-equilibrium phenomenon. There are different strategies existing in the literature to quantify decoherence in the non-equilibrium quantum field theory, e.g. \cite{Calzetta Hu, NEQFT, DTLN, noise, buyanovsky, FCL}.  Usually one computes the decoherence of the system in terms of the von Neumann entropy, e.g. \cite{buyanovsky, buyanovsky1, SHN:2020,  SN:2021, FA, entropy, entropy1, Berges, Schmidt} or the master equation approach \cite{Shaisultanov:1995cf, mastereq, mastereq1, master3}. In this work, we shall instead be interested in an approach based upon the correlation function in order to compute the decoherence in terms of the von Neumann entropy~\cite{Friedrich:2019hev, JFKTPMGS, koksma, kok}. This approach, as usual, divides the entire theory into a system we observe and an environment. The information about the system is characterised by the correlation functions. If one knows all the correlations, one has complete knowledge of the system. However, in a realistic scenario, one expects that only the first few correlation functions and their quantum corrections   (such as the two point, three point or four point functions etc) can be observed. This lack of knowledge about the system then gives rise to an entropy and decoherence. We shall be more specific  about the relevant methodology in the due course.

\begin{figure}
     \centering
     \includegraphics[scale=.50]{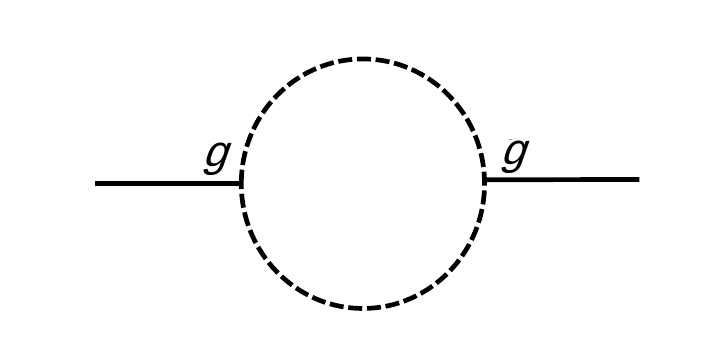}
     %\vspace{-40mm}
     \caption{\small \it One loop self energy diagram for the scalar field for the Yukawa interaction. The dashed lines represent the fermion propagator. This diagram requires scalar field strength renormalisation, as well as a conformal counterterm. See main text for discussion.  }
     \label{figa}
 \end{figure}

We consider here an open quantum field theory  comprising of a massless, minimally coupled  scalar and fermions with the Yukawa interaction  in the inflationary de Sitter spacetime background. We treat the scalar field as the system we observe and the fermions as the surrounding or environment.  We also assume that only the Gaussian 2-point correlation function  are observed by the observer. In this setup, we wish to compute the perturbative von Neumann entropy for the scalar, as a measure of the decoherence due to the interaction with the fermions. We shall restrict ourselves to the one loop self energy for the scalar.  Note that the one loop scalar self energy in the Yukawa theory involves a closed fermion loop~\ref{figa}, and hence no secular effect  is possible at this level of perturbation theory. However, we note also that we are interested to compute the correlator for the scalar field -- and not simply the self energy. The two point correlator corresponding  to the amputated diagram \ref{figa} involves adding external scalar lines, eventually indicating a secular growth for the scalar two point function and its subsequent effect in the decoherence.  The present study is similar to the recent one made in \cite{Bhattacharya:2022wpe}, which considered the same theory in the Minkowski spacetime. Previous literature has considered various types of interactions, such as self-interactions of the inflaton field \cite{Lombardo:2005iz, Martineau:2006ki, Nelson:2016kjm, Nelson:2017pmc}, interactions with gravitational waves \cite{Calzetta:1995ys, piao}, interactions with the system scalar with another scalar~\cite{Rostami:2017akw, Liu:2016aaf, Martin:2018zbe, Martin:2018lin}, as well as interactions with massless and massive fermionic fields \cite{Duffy:2005ue, Prokopec:2003qd, Miao:2006pn, Toms:2018oal, Toms:2018wpy}. We also refer our reader to~\cite{Nusseler:2019ghw, Enqvist:2004pr, Anirban, Lankinen:2019vgv} for some earlier analysis on open quantum systems with scalars and fermions and \cite{Schaub:2023scu, Pethybridge:2021rwf} for the analytical aspects of correlators involving Dirac spinors in de Sitter spacetime with applications to cosmological perturbation theory. The methodology we adopt in this paper will be chiefly based upon~\cite{Friedrich:2019hev,  koksma, kok} and references therein. See also~\cite{Boyanovsky:2018soy} for a study of entropy generation in the Yukawa theory in the de Sitter background via the influence functional technique.

The rest of the paper is themed as follows. \ref{basic} provides a description of the model and includes the formula for the perturbatively corrected phase space area and entropy for the scalar field. \ref{2PF} derives the Kadanoff-Baym equations which are the causal equations of motion for the loop corrected two-point functions in an interacting quantum field theory with self energy corrections, and as well as the statistical propagator. The latter is related to the phase space area necessary to compute the von Neumann entropy.  In \ref{effective}, the two loop two particle irreducible effective action is outlined in the Schwinger-Keldysh or in-in  formalism, which is necessary to compute the  Kadanoff-Baym equations.  \ref{renorm} outlines the derivation of  the renormalised self energy, and the next subsection takes the Fourier transform of it. \ref{statis} finds the perturbative solution for the statistical propagator, phase space area, and the von Neumann entropy of the system, which is a quantifier of the decoherence generated due to  the interaction with the surroundings. The variations of the phase space area and entropy with the relevant parameters are investigated. Finally, \ref{discus} concludes this work. Several necessary computational details are kept in the appendices. 

We shall work in natural units in which $c=\hbar=1$ and with a mostly positive signature of the metric, in $d=4-\e$ dimensions ($\e=0^+$).

%%%%%%%%%%%%%%%
\section{The basic setup} \label{basic}
%%%%%%%%%%%%

The metric for the inflationary de Sitter spacetime, which is our interest,    reads
\begin{eqnarray}
 ds^2 = -dt^2 + e^{2Ht}d\vec{x}\cdot  d\vec{x}   
\end{eqnarray}
where  $H=\sqrt{\Lambda/3}$ is the Hubble constant, and $\Lambda>0$ is the cosmological constant. The above metric written in the conformally flat form reads
\begin{eqnarray}
ds^2 = a^2(\eta) \left[-d\eta^2 + d\vec{x}\cdot  d\vec{x}\right],
\label{l0}
\end{eqnarray}
where $a(\eta)= - 1/H\eta $ is the scale factor and $\eta = - e^{-H t}/H$ is the conformal time. We have the range $0\leq t <\infty$, so that $-H^{-1}\leq \eta < 0^-$.

The action for a massless minimally coupled scalar and fermion coupled via the Yukawa interaction  reads, 
\begin{equation}
S =  \int d^d x a^d\left(-\frac12 (\nabla_\mu\phi)(\nabla^\mu\phi)   - i{\bar{\psi}}\gamma^{\mu}\nabla_{\mu}\psi- g \bar{\psi}\psi \phi \right)
\label{action for 2 fields:dS}
\end{equation}
Note that since we are working with mostly positive signature of the metric, the appropriate anti-commutation relation for the $\gamma$-matrices would be 
$$[\gamma^{\mu},\gamma^{\nu}]_+=-2g^{\mu\nu}{\bf I}_{d\times d}$$
As we have stated in the preceding section, the scalar $\phi(x)$ in \ref{action for 2 fields:dS} will play the role of the system, whereas the fermions will be the environment.   We assume that  the environment is   in its vacuum state. 

Owing to the spatial translational invariance of the de Sitter spacetime,  one can employ the $3$-momentum space. For the free scalar field satisfying $\Box \phi=0$, the temporal part of the modes then reads
\begin{eqnarray}\label{modfunc}
u(\eta, k) =  \frac{H}{\sqrt{2k^3}} (1+ i k \eta)e^{- i k \eta}    
\end{eqnarray}
along with  its complex conjugation, $u^{\star}$. Here $k=|\vec{k}|$ and $u\,(u^{\star})$ respectively corresponds to the positive (negative) frequency modes in the asymptotic past. The above modes define the Bunch-Davies vacuum state, say $|0\rangle$. 
 
 In terms of the asymptotic positive and negative frequency mode functions \ref{modfunc}, it is easy to write down a  $3$-momentum space version of the relevant propagators or two point functions, reviewed in \ref{Propagators in the Schwinger-Keldysh Formalism}. For instance for the Wightman functions, we have
\begin{align}
i \Delta_{\phi}^{\mp \pm}(\eta, \eta^{\prime},k) =  \frac{H^2}{2 k^3}(1\pm i k \eta) (1 \mp i k \eta^{\prime})e^{\mp i k (\eta-\eta^{\prime} )}  \label{BD2PF}
\end{align}

For our present purpose, apart from the usual propagators we require two more kind of two point functions, i.e.  the spectral or the causal two-point function $ \Delta^c_{\phi}$, and the Hadamard or  the statistical two-point function $F_{\phi }$. Their relationship with the Wightman functions and some other properties has been  reviewed in \ref{Propagators in the Schwinger-Keldysh Formalism}. 

From \ref{modfunc}, \ref{defintion of scalar correlators:2}, it is straightforward to compute
\begin{eqnarray}\label{BDDelta} 
  \Delta^c_{\phi}(\eta, \eta^{\prime},k) =  \frac{H^2}{ k^3} \Big[k(\eta-\eta^{\prime}) \cos  k (\eta - \eta^{\prime}) -(1 + k^2 \eta \eta^{\prime}) \sin  k (\eta - \eta^{\prime} ) \Big]   
\end{eqnarray}
and
\begin{eqnarray}\label{BDF} 
 F_{\phi}(\eta, \eta^{\prime},k) = \frac{H^2}{2 k^3} \Big[ (1 + k^2 \eta \eta^{\prime})\cos  k (\eta - \eta^{\prime}) +k(\eta-\eta^{\prime}) \sin  k (\eta - \eta^{\prime} )  \Big]    
\end{eqnarray}
 For our purpose, we need to compute the one-loop correction to the statistical propagator for the Yukawa interaction in the super-Hubble limit.

We note that in analogy with the flat spacetime, the statistical propagator is a mathematical representation of the probability distribution of states, and it describes how the probability density evolves with time. The connection between the statistical propagator and the phase space area arises from the conservation of the phase space volume (i.e., the  Liouville theorem) \cite{JFKTPMGS}. Liouville's theorem states that the phase space volume occupied by a system remains constant during its evolution if the system is isolated. As the system evolves, the statistical propagator governs how the probability density changes, effectively redistributing the same in the phase space. For an open quantum system however, the phase space volume does not remain constant due to the system-environment interactions.

The Gaussian invariant in the momentum space, say $\Xi^2_{\phi}(\eta,k)$, corresponding to the scalar  is given by \cite{Friedrich:2019hev, JFKTPMGS},
\begin{eqnarray}\label{gaussian invariant}
\frac{\Xi_{ \phi }^2(\eta, k)}{4a^4} =  \left[F_{\phi}(\eta, \eta^{\prime},k) \partial_{\eta}\partial_{\eta^{\prime}} F_{\phi}(\eta, \eta^{\prime},k) - \left( \partial_{\eta^{\prime}} F_{\phi}(\eta, \eta^{\prime},k) \right)^2 \right]_{\eta=\eta^{\prime} }   
\end{eqnarray}

Note that $\Xi_{ \phi }^2(\eta, k)$  is dimensionless. The Gaussian part of the von Neumann entropy is given by \cite{Friedrich:2019hev, JFKTPMGS, koksma, kok}
\begin{eqnarray}
 S_{\phi}(\eta, k)  = \frac{\Xi_{ \mathcal{\mathcal{\phi}} } +1}{2} \ln\frac{\Xi_{ \mathcal{\mathcal{\phi}} } +1}{2} -\frac{\Xi_{ \mathcal{\mathcal{\phi}} } -1}{2} \ln \frac{\Xi_{ \mathcal{\mathcal{\phi}} } -1}{2}
 \label{entropyPhi}    
\end{eqnarray}
It is easy to see from \ref{gaussian invariant} using the Bunch-Davies mode functions that for the free case  $\Xi_{ \phi }(\eta, k)$ becomes unity, making the von Neumann entropy written above  vanishing.  This could be attributed to the fact that for the free theory vacuum such as the Bunch-Davies, the uncertainty  is minimal. However, when interactions are introduced, the entropy undergoes perturbative corrections, due to the opening up of new phase space areas. Taking the first order variation of   \ref{gaussian invariant} and using \ref{BDF} gives 
\begin{multline}\label{phase}
\delta \left(\frac{ \Xi^2_{\phi}}{4 a^4} \right)%= \delta \Big[ F_{\phi}(\eta, \eta) \partial_{\eta} \partial_{\eta^{\prime}} F_{\phi}(\eta ,\eta^{\prime}) -  \big[ \partial_{\eta^{\prime}}  F_{\phi}(\eta ,\eta^{\prime})\big]^2 \Big]_{\eta^{\prime} = \eta}
= \Big[  F_{\phi}(\eta, \eta, k) \partial_{\eta} \partial_{\eta^{\prime}} \delta F_{\phi}(\eta ,\eta^{\prime}, k) + \delta F_{\phi}(\eta, \eta, k) \partial_{\eta} \partial_{\eta^{\prime}}  F_{\phi}(\eta ,\eta^{\prime}, k) -2  \partial_{\eta^{\prime}}   F_{\phi}(\eta ,\eta^{\prime}, k)\partial_{\eta^{\prime}}  \delta F_{\phi}(\eta ,\eta^{\prime}, k)\Big]_{\eta=\eta^{\prime}} \\
=   \frac{H^2}{2 k} \Big[\frac{(1 + k^2 \eta^2)}{k^2} \partial_{\eta} \partial_{\eta^{\prime}} \delta F_{\phi}(\eta ,\eta^{\prime}, k) + k^2 \eta^2 \delta F_{\phi}(\eta, \eta, k) - 2   \eta\partial_{\eta^{\prime}}   \delta F_{\phi}(\eta ,\eta^{\prime}, k)\Big]_{ \eta=\eta^{\prime}}\,
\end{multline}

We shall use below the above relation to compute the entropy generation due to the Yukawa interaction. We shall restrict ourselves to one loop order only. We shall use in-in or the Schwinger-Keldysh formalism to compute the various two-point functions.
% \vskip 1cm

%%%%%%%%%%
\section{Kadanoff-Baym equation and the statistical propagator \label{2PF}}
\subsection{The $2$PI effective action}\label{effective}
%%%%%%%%%%

We wish to  compute below  the one  loop correction to the statistical propagator for  the scalar field,  \ref{defintion of scalar correlators:2}, \ref{Propagators in the Schwinger-Keldysh Formalism}, using the in-in or the Schwinger-Keldysh formalism. In order to find out the two point correlation function such as the statistical propagator, it is often useful to construct an equation of motion satisfied by it. Such one particle irreducible correlation functions are generated by a  two particle irreducible (2PI) effective action of the two point function \cite{jackiw}. The equation of motion for the same follows from the first order variation of the effective  action.   Since we wish to look into the one loop effect at the two point functions, we need to construct  a two loop 2PI effective action. For our theory~\ref{action for 2 fields:dS}, the two loop $2$PI effective action in the Schwinger-Keldysh formalism  reads
\begin{figure}[t!]
  \centering
  %\begin{minipage}[t]{.80\textwidth}
   \includegraphics[width=12cm]{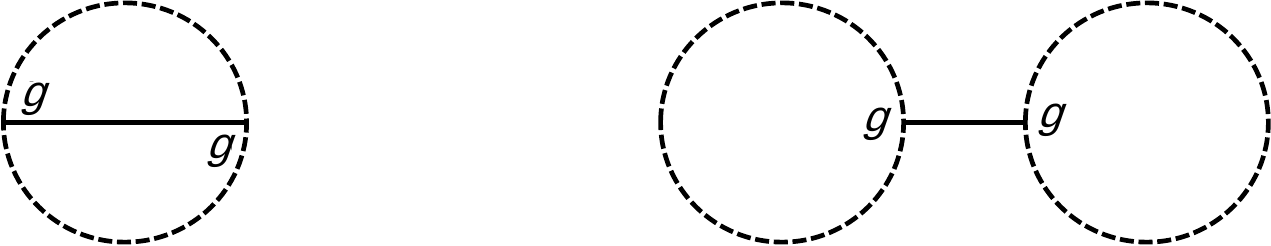}
   \vspace*{0mm}
   \caption{\small \it Contributions to the 2PI effective action due to the Yukawa interaction. The solid line denotes the scalar propagator, whereas the dashed line correspond to the fermion propagator. The Yukawa coupling parameter is denoted by $g$. We  shall restrict our computations to  ${\cal O}(g^2)$ only. The second diagram contains tadpoles and it will not make any contribution. \label{fig:2PIEfAction}}
 %  \end{minipage}
 \end{figure}
\begin{equation}
\Gamma[i\Delta^{ss^{\prime}}_{\phi},i S^{ss^{\prime}}_{\psi}]
 =\Gamma^{(0)}[i\Delta_{\phi}^{ss^{\prime}},i S^{ss^\prime}_{\psi}) ]+\Gamma^{(1)} [i\Delta_{\phi}^{ss^{\prime}},iS^{ss^\prime}_{\psi} ]+\Gamma^{(2)}[i\Delta_{\phi}^{ss^{\prime}},i S^{ss^\prime}_{\psi} ], \quad s ,s^{\prime} = \pm,
 \label{effac}
\end{equation}

where $i\Delta^{ss^{\prime}}_{\phi}$ are the four scalar propagators, whereas $iS^{ss^\prime}_{\psi}$ are the four fermion propagators. The three constituent functionals appearing in \ref{effac} are given by
\begin{align}
\Gamma^{(0)}[i\Delta_{\phi}^{ss^{\prime}},i S^{ss^\prime}_{\psi} ] =& \int d^dxd^dx' a^d \bigg(\nonumber
\sum_{s,s^{\prime}=\pm} \Box_x\delta^d(x-x')\frac{s \,\delta^{ss^\prime}}{2}
i\Delta^{s^{\prime}s}_{\phi}(x',x)\\
&-\sum_{s,s^{\prime}=\pm}i \slashed\nabla_x \delta^d(x-x')s\delta^{ss^\prime}
i S^{s^{\prime}s}_{\psi} (x,x')\bigg)\label{effac0}\\
\Gamma^{(1)} [i\Delta_{\phi}^{ss^{\prime}},iS^{ss^\prime}_{\psi} ]
=&-\frac{i}{2}{\rm Tr}\ln \Big[i \Delta^{ss}_{\phi}(x;x')\Big]+{i}{\rm Tr} \ln \Big[i S^{ss}_{\psi} (x,x) \Big]\label{effac1}\\
\Gamma^{(2)}[i\Delta_{\phi}^{ss^{\prime}},i S^{ss^\prime}_{\psi} ] =&-\sum_{s,s^{\prime}=\pm}\frac{i ss^{\prime}
g^{2}}{2}\int d^dxd^dx' a^d a^{{\prime}d}{\rm Tr}\Big[i S^{ss'}_\psi(x,x')i S^{s's}_\psi(x',x)\Big]
i\Delta^{ss^{\prime}}_{\phi}(x,x')
\label{effac2}
\end{align}

Applying now the variational principle to \ref{effac0}, \ref{effac1} and \ref{effac2} with respect to the scalar propagators,  we obtain the following four equations of motion, i.e. the one loop Kadanoff-Baym equations, 
\begin{align}
\Box_x i\Delta^{ss^{\prime}}_{\phi}(x,x^{\prime \prime})=\frac{s \,  \delta^{ss^{\prime}}i \delta^d(x-x^{\prime \prime})}{a^d}\label{geneom}
+\sum_{s^{\prime\prime}=\pm} \int d^dx^{\prime } a^{{\prime}d}  s^{\prime\prime} \,i  M^{ss^{\prime\prime}}_{\phi}(x,x^{\prime })i\Delta^{s^{\prime\prime}s^{\prime}}_{\phi}(x^{\prime },x^{\prime \prime})
\end{align}
 Thus  the Kadanoff-Baym equations are basically the second order differential equations satisfied by the two point correlators  containing the self energy correction, where the one loop scalar self energies read
\begin{equation}
(aa')^{d}iM^{ss^{\prime}}_{\phi}(x,x^{\prime})=i (aa')^d g^{2}{\rm Tr}\Big[iS^{ss^{\prime}}_\psi(x,x')iS^{s's}_\psi(x',x)\Big]\qquad ({\rm no~sum~on}~s\,{\rm or}\,s')
\label{selfMassIni}
\end{equation}
The corresponding Feynman diagram is given in \ref{figa}. Thus in order to compute the one loop correction to the  statistical propagator defined in \ref{defintion of scalar correlators:2}, we must determine the renormalised self energy first, which we do below. 

%%%%%%%%%%%%
\subsection{The renormalised scalar self energy}\label{renorm}
%%%%%%%%%%%%%%
The renormalised one loop scalar self energy for the Yukawa interaction, \ref{selfMassIni}, was first computed in \cite{Duffy:2005ue}. Since we shall express our final result in a  little bit different form and as well as eventually use the three momentum space as of \cite{Friedrich:2019hev},   we wish to  briefly keep the relevant computations here.  

Let us first express the propagators for the massless fermion field in the de Sitter spacetime, using the de Sitter invariant and complexified length functions
\begin{align}
y_{ss^{\prime}}(x,x') = a(\eta) a(\eta^{\prime} )  H^2  \Delta x_{ss^{\prime}}^2\big(x,x'\big) 
%= \frac{\Delta x_{ss^{\prime}}^2\big(\eta\!-\!\eta^{\prime},\vec x \!-\!\vec x^{\,\prime}\big)  }{\eta \eta^{\prime}}\, 
\label{distance}
\end{align}
where $s,s'=\pm$ as earlier and 
\begin{align}
\Delta x^2_{\pm \pm} &= - \big(| \eta \!-\! \eta^{\prime} | \mp i \varepsilon \big)^2 + | \vec{x} \!-\! \vec{x}^{\,\prime} |^2\, \label{distance1} \\
\Delta x^2_{\pm \mp} &= - \big(\eta \!-\! \eta^{\prime}  \pm i \varepsilon \big)^2 + | \vec{x} \!-\! \vec{x}^{\,\prime} |^2\, \qquad \qquad (\varepsilon=0^+)
\label{distance2}
\end{align}
are the invariant distance functions in the Minkowski spacetime, owing to the conformal flatness of the de Sitter spacetime. 

Since a massless fermion is conformally invariant and the de Sitter spacetime is conformally flat, the fermion propagator  $i {}S^{ss^\prime}_{\psi}(x,x')$ in this background is obtained  by simply acting $i\slashed{\partial}$ on the massless scalar field propagator in the flat spacetime, followed by an overall multiplication by an appropriate power of $a(\eta) a(\eta')$,  
\begin{equation}\label{Feynmanpropposition2}
\begin{aligned}
i {}S^{ss^\prime}_{\psi}(x,x') = (aa^{\prime})^{\frac{1-d}{2}} i\slashed{\partial}\left[\frac{\Gamma\left(\frac{d}{2}-1\right)}{4 \pi^{\frac{d}{2}}}\left[\Delta x_{ss^\prime}^{2}\left(x, x^{\prime}\right)\right]^{1-\frac{d}{2}}\right]
= - \frac{i(aa^{\prime})^{\frac{1-d}{2}} \Gamma\left(\frac{d}{2}\right)}{2 \pi^{\frac{d}{2}}} \frac{\gamma^{\mu} \Delta x_{\mu}}{\left[\Delta x_{ss^\prime}^{2}\left(x, x^{\prime}\right)\right]^{\frac{d}{2}}}
\end{aligned}
\end{equation}
where we have abbreviated, $a(\eta)\equiv a$ and $a(\eta')\equiv a'$. Using the anti-commutation relations for the $\gamma$-matrices appropriate for the mostly positive signature of the metric,  the one loop scalar self energy is then readily found from  \ref{selfMassIni}, \ref{Feynmanpropposition2}
\begin{equation}\label{SelfMassPosspace}
i (aa')^d M_{\phi}^{++}(x,x')= - 
(aa')^d\frac{i g^{2}(aa^{\prime})^{{1-d}} \Gamma^{2}(\frac{d}{2})}{ \pi^{\scriptscriptstyle{d}}}
\frac{1}{ \Delta x_{++}^{2d-2}(x,x')}
\end{equation}
Similarly, we can find out the other self-energies $ (aa')^di M_{\phi}^{--}(x,x')$, $(aa')^di M_{\phi}^{+-}(x,x')$ and $(aa')^d i M_{\phi}^{-+}(x,x')$  using
the suitable $i\epsilon$ prescriptions. Using now \ref{distance}, we obtain
\begin{equation}\label{SelfMassPo}
i M_{\phi}^{++}(x,x')= - 
\frac{i g^{2}\Gamma^{2}(\frac{d}{2}) H^{2d-2}}{ \pi^{\scriptscriptstyle{d}}} y_{++}^{1-d}
\end{equation}
Using next the identity for $y^{1-d}$ given in \ref{defAndConv}, we rewrite $i M_{\phi}^{++}(x,x')$ as
\begin{eqnarray}
i M_{\phi}^{++}(x,x')= -
\frac{i g^{2}\Gamma^{2}(\frac{d}{2}) H^{2d-2}}{ 2^{2d-2}\pi^{\scriptscriptstyle{d}}} \Bigg[\frac{2}{(d-2)^2} \frac{\square}{H^2} -\frac{2}{(d-2)}\Bigg] \Big( {\frac{y_{++}}{4}}\Big)^{2-{d}} 
\end{eqnarray}
where the dimensionless d'Alembertian  operator in the de Sitter spacetime reads
\begin{align}
\frac{\square}{H^2} = \eta^2 \Big[- \partial^2_{\eta} + \frac{d-2}{\eta} \partial_{\eta} + \vec{\partial}^2  \Big] 
\end{align}
Substituting the value of $\Big( {\frac{y_{++}}{4}}\Big)^{2-{d}} $ from \ref{defAndConv}, we now obtain
\begin{multline}
i M_{\phi}^{++}(x,x') = \frac{i g^{2}\Gamma^{2}(\frac{d}{2}) H^{2d-2}}{ 2^{2d-2}\pi^{\scriptscriptstyle{d}}} \Bigg[\frac{2}{(d-2)^2} \frac{\square}{H^2} -\frac{2}{(d-2)}\Bigg]  \Bigg[  \frac{\square}{H^2} \Big(\frac{4}{y_{++}}\ln \frac{\mu^2 y_{++}}{H^2}  \Big)   - \frac{4}{y_{++}} \Big(2 \ln \frac{\mu^2 y_{++} }{H^2}  -1\Big) \Bigg]   
 + \mathcal{O} \big( d-4 \big) 
\label{some equation}
\end{multline} 
where $\mu$ is renormalisation scale with mass dimension one. Apart from the ultraviolet finite terms, \ref{some equation} yields a divergent local contribution to the self energy  
\begin{equation}
(aa')^d i M^{ss^{\prime}}_{\phi}(x,x')\vert_{\rm div}
  = -
\frac{i (aa')^dg^{2}\Gamma^{2}(\frac{d}{2}) H^{2d-2}}{ 2^{2d-2}\pi^{\scriptscriptstyle{d}}} \Bigg[\frac{2}{(d-2)^2} \frac{\square}{H^2} -\frac{2}{(d-2)}\Bigg]  \Bigg[ \frac{2 (4 \pi )^{d/2} }{(d-3)(d-4)\Gamma\big[\frac{d}{2}-1\big] } \Big(  \frac{ \mu}{H}\Big)^{d-4}  \frac{i \delta^d \big(x-x^{\prime} \big)}{(Ha)^d} \Bigg]
  s \delta^{ss^{\prime}}
\label{counterterm self energy}
\end{equation}
As was shown in~\cite{Duffy:2005ue}, this divergence can be absorbed by the scalar field strength renormalisation  {\it and} a conformal counterterm, such that their combination introduces a term in the Lagrangian density
$$  (aa')^{d/2} \left( \frac{\delta Z}{aa'} \p^2 \delta^d (x-x')\right)  $$
whose contribution, when added to the self energy, leads  us to the choice of the scalar field strength renormalisation counterterm 
\be
\delta Z = -\frac{ g^2 \mu^{-\e} \Gamma(1-\e/2) }{2^2\pi^{2-\e/2}\e(1-\e)}
\label{y13}
\ee
  The resulting one loop self energy after renormalisation reads
\begin{eqnarray}
(aa')^d i M^{++}_{\phi, \text{ren}}(x,x') =  \frac{i g^{2} H^6 (aa')^d}{ 2^6\pi^4} \Bigg[\frac{\square}{2H^2} -1\Bigg]  \Bigg[  \frac{\square}{H^2} \Big(\frac{4}{y_{++}}\ln \frac{\mu^2 y_{++}}{H^2} \Big) - \frac{4}{y_{++}} \Big(2 \ln \frac{\mu^2 y_{++} }{H^2}  -1\Big) \Bigg]  
\label{renormalized self mass}   
\end{eqnarray}
The other renormalised self energies, $i M^{ss^{\prime}}_{\phi, \text{ren}}(x,x')$ ($s,s^{\prime}=\pm$), can be found in a similar manner. However,  the self-energies of mixed kinds, i.e. those associated with the Wightman functions are free of divergences and hence do not require any renormalisation. 

%%%%%%%%
\subsection{Self energy in the  spatial momentum space}
%%%%%%%%%

 Following \cite{Friedrich:2019hev}, we use the identities \ref{i1}, \ref{i2} derived in \ref{defAndConv}  to rewrite  \ref{renormalized self mass} as,
\begin{multline}
\label{inhomSelfMass}
i M^{++}_{\phi, \text{ren}}(x, x^{\prime}) = \frac{i g^{2} H^6}{ 2^6\pi^4} \Bigg[ \frac{\square}{2H^2} -1\Bigg] \Bigg\lbrace \frac{1}{4}
\frac{\square^2}{H^4} \Bigg[  \frac{1}{2}\ln^2\frac{y_{++}}{4}+   \ln  \frac{4 \mu^2 }{ e H^2}  \ln \frac{y_{++}}{4} \Bigg]	
\\ -  \frac{1}{2}  \frac{\square}{H^2} \Bigg[ \frac{1}{2}\ln^2 \frac{y_{++}}{4}
+  \ln  \frac{4 \mu^2 }{ e^3 H^2}  \ln  \frac{y_{++}}{4} \Bigg] - \frac{3}{2} \ln \frac{ y_{++}}{4} 
\Bigg\rbrace %+ \text{hom}
\end{multline}
which we rewrite as 
\begin{multline}
\label{inhomSelfMass1}
i M^{++}_{\phi, \text{ren}}(x, x^{\prime}) = \frac{i g^{2} H^6}{ 2^7\pi^4}  \Bigg\lbrace \frac{1}{4}
\Bigg(\frac{\square}{H^2}\Bigg)^3 \Bigg[  \frac{1}{2}\ln^2 \frac{y_{++}}{4} + \ln \frac{4 \mu^2 }{ e H^2}  \ln \frac{y_{++}}{4} \Bigg]	
 -   \Bigg(\frac{\square}{H^2} \Bigg)^2 \Bigg[ \frac{1}{2}\ln^2 \frac{y_{++}}{4}
+  \ln  \frac{4 \mu^2 }{ e^2 H^2}  \ln  \frac{y_{++}}{4} \Bigg] \\
+ \Bigg(\frac{\square}{H^2} \Bigg) \Bigg[ \frac{1}{2}\ln^2 \frac{y_{++}}{4}
+  \ln  \frac{4 \mu^2 }{ e^{\frac{9}{2}} H^2}  \ln \frac{y_{++}}{4} \Bigg] 
- 3 \ln \frac{ y_{++}}{4} 
\Bigg\rbrace 
\end{multline}
We now take the spatial Fourier transform of \ref{inhomSelfMass1} defined by  
\begin{align}
i M^{++}_{\phi, \text{ren}}\big(\eta, \eta^{\prime} ,k \big) = \int d^{3}\vec{r}\, i M^{++}_{\phi, \text{ren}}\big(x, x^{\prime} \big) e^{-i \vec{k} \cdot\vec{r}}
\end{align}
and use \ref{fTlogsApp}, derived in \ref{fourier}, in order to find  in the momentum space
\begin{multline}
i M^{++}_{\phi, \text{ren}} \big(\eta, \eta^{\prime},k \big)  =  \frac{i g^{2} H^6}{ 2^7\pi^4} \Bigg( -\frac{4 \pi^2}{k^3} \Bigg)\Bigg\lbrace \frac{1}{4}\Bigg(\frac{\square_k}{H^2}\Bigg)^3 \Bigg( \Bigg[2+ \big[1+ i k |\Delta \eta | \big] \Big(  \ln \frac{ 2 | \Delta \eta| \mu^2}{e k \eta \eta^{\prime} H^2 }+  \frac{i \pi}{2}- \gamma_E \Big) \Bigg] e^{-i k |\Delta \eta|} \\ 
- \big(1 - i k |\Delta \eta| \big)\Bigg[ \text{ci} \big[ 2 k| \Delta \eta|  \big]  -i \,  \text{si} \big[ 2 k |\Delta \eta|  \big]  \Bigg] e^{+i k |\Delta \eta|} \Bigg) \\
 - \Bigg(\frac{\square_k}{H^2}\Bigg)^2 \Bigg(\Bigg[ 2+ \big[1+ i k |\Delta \eta | \big] \Big(  \ln \frac{ | \Delta \eta|H^2}{ e^2 k \eta \eta^{\prime} \mu^2} +  \frac{i \pi}{2}- \gamma_E \Big)  \Bigg]   e^{-i k |\Delta \eta|} 
- \big(1 - i k |\Delta \eta| \big)\Bigg[ \text{ci} \big[ 2 k| \Delta \eta|  \big]  -i \,  \text{si} \big[ 2 k |\Delta \eta|  \big]  \Bigg] e^{+i k |\Delta \eta|}  \Bigg)\\
+ \Bigg(\frac{\square_k}{H^2}\Bigg) \Bigg(\Bigg[ 2+ \big[1+ i k |\Delta \eta | \big] \Big(  \ln \frac{ | \Delta \eta|H^2}{ e^{\frac{9}{2}} k \eta \eta^{\prime} \mu^2} +  \frac{ i\pi}{2}- \gamma_E \Big)  \Bigg]   e^{-i k |\Delta \eta|} \\
- \big(1 - i k |\Delta \eta| \big)\Bigg[ \text{ci} \big[ 2 k| \Delta \eta|  \big]  -i \,  \text{si} \big[ 2 k |\Delta \eta|  \big]  \Bigg] e^{+i k |\Delta \eta|}  \Bigg)
+3 \Bigg[\big[1+ i k |\Delta \eta | \big]  e^{-i k |\Delta \eta|} \Bigg] \Bigg\rbrace
  \label{M++Full}
\end{multline}
where 
$\square_k$  is the d'Alembertian in the momentum space,
$$\square_k = \eta^2 \Big[- \partial^2_{\eta} + \frac{d-2}{\eta} \partial_{\eta} -\vec{k}^2  \Big] $$
and  $\rm ci$, $\rm si$ are the cosine and sine integral functions respectively, given at the beginning of  \ref{defAndConv}.
 \ref{M++Full} seems to be involved enough to be handled  in the Kadanoff-Baym equations, \ref{geneom}. However, since our chief interest is to look into the dynamics of the quantum field at late times or towards the end of inflation, we shall take its super-Hubble or the infrared limit, i.e., $\eta, \eta' \to 0$. In this limit \ref{M++Full} simplifies to  
\begin{eqnarray}
i M^{++}_{\phi, \text{ren}} (\eta, \eta^{\prime},k)_{ k | \Delta \eta| \ll1}  \approx  \frac{i g^{2} H^6}{ 2^5\pi^2 k^3} \Bigg[ \frac{1}{4}\Bigg(\frac{\square}{H^2}\Bigg)^3 
 -\Bigg(\frac{\square}{H^2}\Bigg)^2 
+\Bigg(\frac{\square}{H^2}\Bigg) \Bigg] \Bigg( \ln \frac{H^2 k^2 \eta \eta^{\prime}}{\mu^2} + 2 i k  |\Delta \eta |  \Bigg)
  \label{M++Full15}
\end{eqnarray}
Note that unlike the case of a massless minimal scalar field \cite{nitin, Bhattacharya:2022aqi}, the above self energy is devoid of any secular logarithm  of the scale factor \cite{Miao:2006pn}, $a=-1/H\eta$. This is not surprising, as the loop we have computed consists of fermion propagators only. 
Using \ref{M++Full15}, we shall now  compute the statistical propagator and the von Neumann entropy in the next section. 

%%%%%%%%%%%
\subsection{Perturbative solution for the statistical propagator and the von Neumann entropy} \label{statis}
%%%%%%%%%%%%

Let us look at the renormalised version of  equations of motion \ref{geneom} for the propagators $i \Delta_{\phi}^{ss^{\prime}}(x, x^{\prime }) $.
By rewriting the two-point functions in terms of real and imaginary parts, we obtain after using \ref{defintion of scalar correlators} into \ref{geneom}
\begin{multline}
\square_{x} F_{\phi} (x,x^{\prime \prime})  = \frac{i}{2} \int  d^4 x^{\prime }\big(\eta^{\prime} H \big)^{-4}\, \Big[    M^{++}_{\phi,\text{ren}}(x, x^{\prime })  -   M^{--}_{\phi,\text{ren}}(x, x^{\prime }) +   M^{-+}_{\phi,\text{ren}}(x, x^{\prime }) -   M^{+-}_{\phi,\text{ren}}(x, x^{\prime })   \Big] F_{\phi} (x^{\prime }, x^{\prime\prime} )   \\ - \frac{1}{4} \int   d^4 x^{\prime }\big(\eta^{\prime} H \big)^{-4}\, \Big[\text{sign} (\eta^{\prime} - \eta^{\prime \prime} ) \Big(    M^{++}_{\phi,\text{ren}}(x, x^{\prime })  +    M^{--}_{\phi,\text{ren}}(x, x^{\prime })\Big) -   M^{-+}_{\phi,\text{ren}}(x, x^{\prime }) -   M^{+-}_{\phi,\text{ren}}(x, x^{\prime })   \Big]  \Delta_{\phi}^c (x^{\prime }, x^{\prime\prime} )  \label{statEQX}
\end{multline}

 We wish to solve the above equation perturbatively for the statistical propagator at ${\cal O}(g^2)$, using the one loop self energy.  
In order to do this, we will  substitute the expressions from the free theories given by \ref{BDDelta}, \ref{BDF} for $\Delta_{\phi}^c$ and $ F_{\phi}$, on the right-hand side of \ref{statEQX}. We have  in the momentum space
\begin{eqnarray}
\square_k F_{\phi} (\eta,\eta^{\prime \prime},k)\approx  \int_{- 1/H}^{\eta} d \eta^{\prime} \big(\eta^{\prime} H \big)^{-4}\, M^c(\eta, \eta^{\prime} ,k) F_{\phi} (\eta^{\prime }, \eta^{\prime\prime},k )  + \int_{-1/H}^{\eta^{\prime \prime}} d \eta^{\prime} \big(\eta^{\prime} H \big)^{-4}\, M^F(\eta, \eta^{\prime },k)  \Delta_{\phi}^c (\eta^{\prime }, \eta^{\prime\prime},k )
\end{eqnarray}
where we have used \ref{energies} and \ref{energies1} given in \ref{defAndConv}. We note that while utilising \ref{energies1} into \ref{statEQX}, it becomes evident that $\eta \gtrsim \eta^{\prime}$ and $\eta^{\prime} \lesssim \eta^{\prime \prime}$, consequently determining the upper limits of the above integration. The lower limit is set to be $-1/H$, as explained in \ref{basic}. Using now \ref{newenergies} and \ref{newenergies1} in above equation, we obtain
\begin{multline}
\square_k F_{\phi} (\eta,\eta^{\prime \prime},k)  \approx -2  \int_{- 1/H}^{\eta} d \eta^{\prime} \big(\eta^{\prime} H \big)^{-4} \text{Im} M^{++}(\eta, \eta^{\prime} ,k) F_{\phi} (\eta^{\prime }, \eta^{\prime\prime},k )\\  - \int^{\eta^{\prime \prime} }_{-1/H} d \eta^{\prime} \big(\eta^{\prime} H \big)^{-4}\,  \text{Re} \,  M^{++}(\eta, \eta^{\prime} ,k) \Delta_{\phi}^c (\eta^{\prime }, \eta^{\prime\prime},k ) \label{eqnF}
 \end{multline}
Since we shall use   the free theory results for the propagators appearing on the right hand side, we may solve the above equation by using a Green function.  The causal result corresponds to the retarded  Green function $G_{\text{ret}}(\eta , \eta^{\prime} , k)$, satisfying in the momentum space  
\begin{align}
\square_k G_{\text{ret}}(\eta , \eta^{\prime} , k) = H^2 \eta^2 \Big[  - \partial_{\eta}^2   + \frac{d-2}{\eta} \partial_{\eta} - k^2 \Big]G_{\text{ret}}(\eta , \eta^{\prime} , k) =  a^{-4}(\eta^{\prime}) \delta(\eta-\eta^{\prime}) 
\end{align}
which can be easily solved by using the free theory mode functions, e.g.~\cite{Friedrich:2019hev},
\begin{eqnarray}
G_{\text{ret}}(\eta , \eta^{\prime} , k) = \theta(\eta - \eta^{\prime}) \frac{H^2}{k^3} \Big[ k(\eta - \eta^{\prime}) \cos  k(\eta - \eta^{\prime} )  -(1 + k^2 \eta \eta^{\prime} ) \sin k(\eta - \eta^{\prime}) \Big],
\end{eqnarray}
which, in the late time or super-Hubble limit we are interested in  becomes
\begin{eqnarray}
G_{\text{ret}}(\eta , \eta^{\prime} , k)_{ k | \Delta \eta| \ll1} = -\theta(\eta-\eta^\prime) \frac{H^2 | \Delta \eta|^3}{3}  
\end{eqnarray}
After substituting the real and imaginary parts of $i M^{++}_{\phi, \text{ren}}$ from \ref{M++Full15} into \ref{eqnF}, we find the one loop corrected statistical propagator in the supper-Hubble limit
\begin{multline}
F_{\phi} (\eta,\eta^{\prime \prime},k)_{ k | \Delta \eta| \ll 1}   =  \frac{1}{H^2}\left[-\frac{1}{4} \Bigg(\frac{\square_k}{H^2} \Bigg)^2 + \Bigg(\frac{\square_k}{H^2} \Bigg)  - 1 \right]\mathcal{O}(\eta, \eta^{\prime \prime},k)_{ k | \Delta \eta| \ll 1} + H \int_{-\infty}^{\infty}d\tau \frac{G_{\text{ret}}(\eta, \tau,k)_{ k | \Delta \eta| \ll 1} }{(\tau H)^4}\\+ F_{\text {free}} (\eta ,\eta^{\prime \prime},k)   \label{eqnFInv}
\end{multline}
where we have abbreviated 
\begin{multline}
\mathcal{O}(\eta, \eta^{\prime \prime},k)_{ k | \Delta \eta| \ll 1}  = - \frac{ g^{2} H^6}{ 2^3\pi^2 k^3}   \int_{- 1/H}^{\eta} d \eta^{\prime} \big(\eta^{\prime} H \big)^{-4} k  |\Delta \eta | F_{\phi} (\eta^{\prime }, \eta^{\prime\prime},k )_{{\rm free}, k | \Delta \eta| \ll 1}  \\- \frac{ g^{2} H^6}{ 2^5\pi^2 k^3} \int^{\eta^{\prime \prime} }_{-1/H} d \eta^{\prime} \big(\eta^{\prime} H \big)^{-4}\ln \frac{H^2 k^2 \eta \eta^{\prime}}{\mu^2}  \Delta_{\phi}^c (\eta^{\prime}, \eta^{\prime\prime},k )_{{\rm free}, k | \Delta \eta| \ll 1}   \label{BLog}
\end{multline}
and
$
\square_k F_{\text{free}} (\eta ,\eta^{\prime \prime},k)=0$ is the free theory part of the statistical propagator, \ref{BDF}. The expressions for  $F_{\phi} (\eta^{\prime }, \eta^{\prime\prime},k )_{ {\rm free}, k | \Delta \eta| \ll 1}$ and $\Delta_{\phi}^c (\eta^{\prime }, \eta^{\prime\prime},k )_{{\rm free}, k | \Delta \eta| \ll 1}$  in the infrared or super-Hubble  limit are explicitly given in \ref{nc4}, \ref{corr1}.  Substituting everything now into \ref{eqnFInv} and integrating, we have the result for the statistical propagator up to one loop
\begin{multline}\label{sss}
F_{\phi} (\eta,\eta^{\prime \prime},k)_{ k | \Delta \eta|\ll 1}\vert_{\rm 1\,loop} \approx \frac{g^{2} H^2}{768 \pi ^2 k^3}\Bigg[ \ln \frac{\eta \eta '' H^2 k^2}{\mu ^2} \left(\left(\eta ^4 k^4+2 \eta ^2 k^2+4\right) \ln \frac{\eta \eta '' H^2 k^2 }{\mu ^2}-4  \left(\eta ^2 k^2+6\right)\right)\\
   +4 \left(\eta ^2 k^2+6\right) \ln \left(-\frac{\eta  H k^2}{\mu ^2}\right)-\left(\eta ^4 k^4+2 \eta ^2 k^2+4\right) \ln
   ^2\left(-\frac{\eta  H k^2}{\mu ^2}\right)\Bigg]  + F_{\text {free}} (\eta ,\eta^{\prime \prime},k)
\end{multline}

However, note that the statistical propagator must be symmetric under the interchange of its coordinate arguments, \ref{statistical}. In other words, while acting $\square^{\prime \prime}_k$ on $F_{\phi}(\eta^{\prime \prime}, \eta, k)$, we must reproduce the right hand side of \ref{eqnF} with $\eta$ and $\eta^{\prime \prime}$ interchanged. This implies the correct expression for the statistical propagator is achieved after symmetrising \ref{sss} with respect to its coordinate arguments~\cite{Friedrich:2019hev}, 

\begin{equation}\label{ssss}
    F_{\phi}(\eta, \eta^{\prime \prime}, k)_{k |\Delta \eta| \ll 1}\Big|_{\rm 1\,loop} = \frac{1}{2}\Big[F_{\text {free}}(\eta, \eta^{\prime \prime}, k) + F_{\text {free}}(\eta^{\prime \prime}, \eta, k)\Big] + \frac{1}{2}\Big[F_{\text {int}}(\eta^{\prime \prime}, \eta, k) + F_{\text {int}}(\eta, \eta^{\prime \prime}, k)\Big]
\end{equation}
We also note the consistency relation for the interacting part of \ref{sss}

\begin{equation}
\square_k \square^{''}_k \Big[F_{\text {int}}(\eta, \eta^{\prime \prime}, k) - F_{\text {int}}(\eta^{\prime \prime}, \eta, k)\Big]  = 0  
\end{equation}

\bigskip

\noindent Let us now substitute \ref{sss} and \ref{ssss} into \ref{phase}, in order to obtain the increase in the phase space area 
\begin{eqnarray}
\delta \left(\frac{ \Xi_\phi ^2}{4a^4} \right) \approx \frac{g^{2} H^4 \left(\eta ^2 k^2 \ln \frac{\eta ^2 H^2 k^2}{\mu ^2} \left(\eta ^2 k^2 \ln \frac{\eta ^2 H^2 k^2}{\mu ^2}-4\right)+2 \eta ^2
   k^2+2\right)}{384 \pi ^2 \eta ^2 k^6}
\end{eqnarray}

\begin{figure}[!ht]
    \centering
    \includegraphics[scale=0.52] {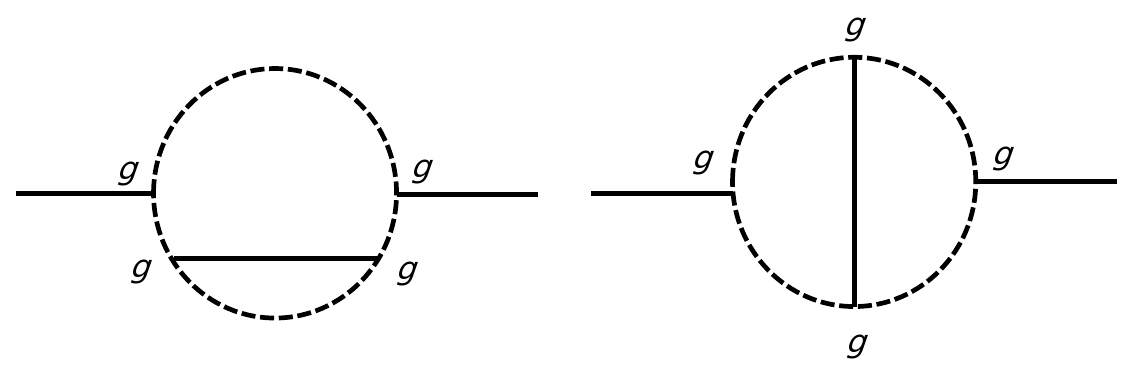}\hspace{1.0cm}
    \caption{\small \it Self energy diagrams for the scalar field at two loop order for the Yukawa coupling. The broaken lines denote fermions whereas the solid lines denote the scalars. See main text for discussion.
 }
    \label{selfattwoloop}
\end{figure}

Converting $\eta$ now in terms of the scale factor $a$, and keeping the late time ($\eta \to 0^-$) dominant terms only, we note in particular
\begin{eqnarray}
\delta \Xi_\phi ^2\approx \frac{ g^{2} H^6 a^6}{48 \pi ^2 k^6} 
\label{areaadd} 
\end{eqnarray}
\begin{figure}[!ht]
    \centering
    \includegraphics[scale=0.55] {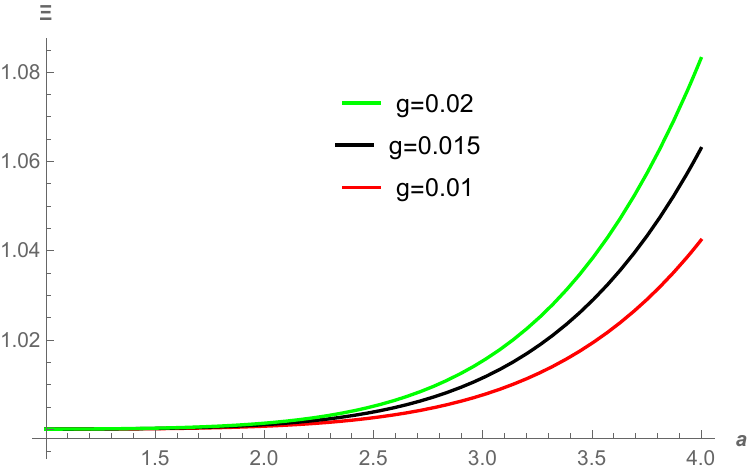}\hspace{1.0cm}
     \includegraphics[scale=0.55] {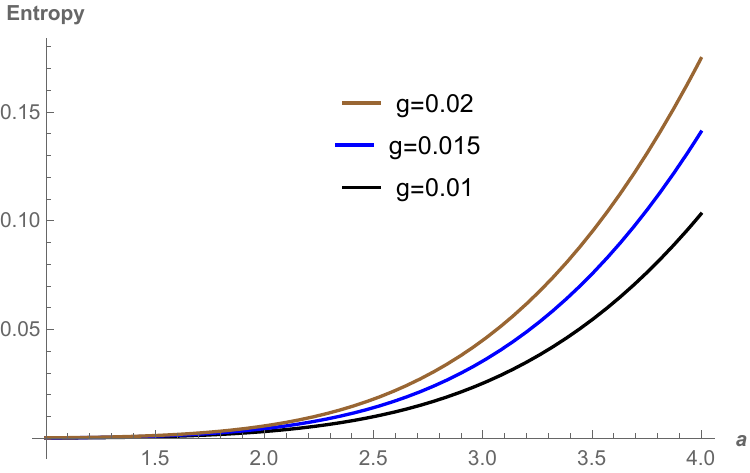}
    \caption{\small \it Variation of the phase space area (left) and the von Neumann entropy (right) with respect to the scale factor. Both  increase monotonically with increasing $a$ and the Yukawa coupling, as expected. }
    \label{fig:phasespace}
\end{figure}
By taking (for example) $k/H\sim1$, we have  plotted the above variation  in \ref{fig:phasespace} for different values of the Yukawa coupling. We next use \ref{entropyPhi} in order to compute the von Neumann entropy,
\begin{equation}
 S_{\phi}(\eta, k)  = \frac{\left(a^6 g^2 H^6+96 \pi ^2 k^6\right) \log \left(\frac{a^6 g^2 H^6}{96 \pi ^2 k^6}+1\right)-a^6 g^2 H^6 \log \left(\frac{a^6 g^2 H^6}{96 \pi ^2 k^6}\right)}{96 \pi ^2 k^6}    
\end{equation}
which is manifestedly a positive quantity and has been depicted in \ref{fig:phasespace}. Expectedly, the von Neumann entropy increases monotonically with increasing scale factor and the coupling constant, and vanishes as $g\to 0$, similar to that of the increment in the phase space area, \ref{areaadd}. Note also that the above result is perturbative and limited to one loop only. Progressing to higher orders in the perturbation theory will definitely bring into more late time non-perturbative secular logarithms originating from the internal scalar propagators residing in the self energy loops, \ref{selfattwoloop}. It seems that, a physically meaningful answer (for $k \ll H a$, with say $g \sim {\cal O}(1)$) can then be achieved only via some appropriate resummation procedure. This seems to be a challenging task, as for example, the two loop  diagrams of \ref{selfattwoloop} are likely to make non-local contributions to the scalar self energy. Verification of these anticipations requires addressing this problem in a non-perturbative framework. This is beyond the scope of the present paper and we wish to reserve it for a future work.

%

%%%%%%%
\section{Discussion \label{discus}}
%%%%%%%%%

In this work, we have studied the decoherence and subsequent generation of entropy for the Yukawa interaction in the inflationary de Sitter spacetime, using the correlator formalism developed in~\cite{Friedrich:2019hev, JFKTPMGS, koksma, kok}. We have treated the scalar as our system, whereas the fermions as the surrounding or environment. This work is  an extension of our previous analysis  in the Minkowski spacetime \cite{Bhattacharya:2022wpe}. Our chief objective was to compute the perturbative von Neumann entropy for the scalar at one loop.  The entropy turns out to be monotonically increasing with respect to the time (or the scale factor), as well as the Yukawa coupling, \ref{fig:phasespace}.  This is in agreement with the case when both the system and the surrounding are scalar fields~\cite{Friedrich:2019hev}.  Our result is also in qualitative  agreement with~\cite{Boyanovsky:2018soy}, where such entropy generation was computed using the Feynman-Vernon influence functional technique.  However, compared to~\cite{Friedrich:2019hev}, we also note a difference in the logarithmic terms of the statistical propagator,~\ref{sss}. Specifically, our statistical propagator appears to be logarithmically less enhanced compared to that of~\cite{Friedrich:2019hev}. This is a qualitative difference, and  corresponds to the fact that a massless fermion loop is conformal and a fermion propagator contains a derivative acting on the scalar propagator. Nevertheless, at sufficiently late times $\eta \to 0^{-}$, the most significant correction to the phase space area, \ref{phase} and \ref{areaadd}, originates from the momentum-momentum correlator, $(\partial_{\eta}\partial_{\eta^{\prime}} F_{\phi}(\eta, \eta^{\prime},k))$, exhibiting a proportionality with $a^6$ in both cases (scalar-scalar and scalar-fermions), up to multiplicative numerical factors. The logarithmic terms in the correction to the phase space area are sub-leading. This yields  the leading late time growth of the von Neumann entropy to be proportional to $\ln a$, indicating a rapid classicalisation of quantum fluctuations, at least as of this perturbative computation. The increase in the entropy with the scale factor naturally corresponds to the fact that the phase space area must increase with time in an expanding spacetime. The increase of the entropy with respect to the Yukawa coupling must correspond to the increase or opening up of new phase space areas due to the interaction. This feature is qualitatively similar to that of the Minkowski spacetime \cite{Bhattacharya:2022wpe}, even though this result in the Minkowski spacetime was non-perturbative.  

The impact of loop corrections for interacting  quantum fields become particularly significant towards the end of the inflation. Note that, in order to leave any significant footprint on the cosmic microwave background, the fields should remain sufficiently light for a certain duration in the radiation dominated  era, e.g.~\cite{Friedrich:2019hev} and references therein. Now, we have considered here a spectator scalar field coupled to fermions in the de Sitter background. What happens when we consider more general quasi-de Sitter background by taking an inflaton? The fermion as well as some spectator field can couple to the inflaton. In order to investigate decoherence in these contexts, perhaps formulating the whole problem stochastically would be useful, for example, by constructing an one loop effective action by integrating out the fermionic degrees of freedom, e.g.~\cite{Miao:2006pn}. Attempting to construct a non-perturbative effective action also seems to be interesting in this context. Perhaps investigating the classicalisation of the primordial quantum gravitational perturbations via interactions with spectator fields/inflaton would be an important task as well. Finally, as we have already mentioned towards the end of the preceding section, the  analysis of this paper is perturbative and at one loop only. Going to higher order of the perturbation theory is expected to bring in the secular logarithms due to internal scalar propagators.  The non-perturbative resummation of these effects using the Kadanoff-Baym equations poses a challenging task. Such non-perturbative analysis seems to be also necessary to fully understand the consequence of the aforementioned qualitative difference of the logarithmic enhancement of the statistical propagator, for the scalar-scalar~\cite{Friedrich:2019hev} and scalar-fermion interactions. We intend to delve into these issues in our future publications.

\bigskip
\section*{Acknowledgement}
The authors would like to sincerely acknowledge anonymous referee for a careful critical reading of an earlier version of this  manuscript, and for making various useful comments.
%%%%%%

\bigskip\bigskip
\bigskip

%\paragraph{Acknowledgments}\vspace{1mm}

%\pagebreak

\section*{Appendices}

\appendix
\labelformat{section}{Appendix #1}

%%%%%%%
\section{The Schwinger-Keldysh formalism and propagators}
\label{Propagators in the Schwinger-Keldysh Formalism}
%%%%%%

The conventional in-out formalism for the $S$-matrix in quantum field theory loses its utility in dynamical  or non-equilibrium  contexts where the initial vacuum state can decay due to particle pair creation. The cosmological spacetime we are interested in serves as such an instance. In these cases, the Schwinger-Keldysh formalism, also known as the in-in or closed time path formalism, furnishes  a very useful  framework to compute {\it causal} expectation values with respect to some initial state, even when the final states are not known explicitly~\cite{Schwinger:1960qe, Keldysh:1964ud}.  Specifically, this formalism involves  the time evolution of an operator starting from an initial state defined at some $\eta =\eta_0$ to some future hypersurface at $\eta=\eta_f$. Utilising the completeness relationship of the states residing on this hypersurface (and without requiring to know any further detail of them), the operator is then evolved backward  in time. In the presence of interactions, it is clear that   the evolution of the operator from $\eta_0$ to $\eta_f$ involves a time-ordering, followed by an anti-time ordered evolution from $\eta_f$ back to some earlier time.  Accordingly in this formalism, we can express the expectation value of an operator $O(\eta)$ with respect to some initial density operator $\rho(\eta_{0})$ as follows
\begin{equation} \label{expectationvalues}
\langle O(\eta) \rangle =
\mathrm{Tr}\left[\rho(\eta_{0})O(\eta)\right] =
\mathrm{Tr}\left[ \rho(\eta_{0}) \left\{ \overline{T}
\exp\left(i \int_{\eta_0}^{\eta_f} a(\eta')\mathrm{d}\eta^\prime 
H(\eta^\prime)\right) \right\} O(\eta_{0}) \left\{ T \exp
\left(-i \int_{\eta_0}^{\eta_f} a(\eta'') \mathrm{d} \eta'' H(\eta'')
\right) \right\}\right]
\end{equation}
 where $ H(\eta)$ is the Hamiltonian and $\overline{T}$ stands for  anti-time  ordering. The corresponding generating functional  for a theory containing scalar and fermions is written as
\begin{eqnarray}\label{Z:inin}
 && {\cal Z}[J_{+}^{\phi}, J_{-}^{\phi}, J_{+}^{\psi},
J_{-}^{\psi}, \rho(\eta_0)]  \!=\! \int \! {\cal D}\phi^{+}_{0}{\cal D}\phi^{-}_{0}
[{\cal D}\psi^{+}_{0}] [{\cal D}\psi^{-}_{0}] \langle\phi^{+}_{0},
\psi^{+}_{0}| { \rho}(\eta_{0})|\phi^{-}_{0}, \psi^{-}_{0}
\rangle \nonumber\\
&& \qquad\!\int_{\phi_{0}^{+}}^{\phi_{0}^{-}}\! {\cal
D}\phi^{+}{\cal D}\phi^{-}
\delta[\phi^{+}(\eta_{f}\!)-\phi^{-}(\eta_{f}\!)]  \!
\int_{\psi_{0}^{+}}^{\psi_{0}^{-}} \![{\cal D}\psi^{+}][{\cal
D}\psi^{-}] [\delta[\psi^{+}(\eta_{f}\!)-\psi^{-}(\eta_{f}\!)]]
\nonumber \\
&& \qquad \times {\rm exp}\left[i \int
\mathrm{d}^{d-1} \vec{x}\int_{\eta_{0}}^{\eta_{f}} a^d(\eta')\mathrm{d}\eta^\prime
\left({\cal L}[\phi^{+},\psi^{+}]-{\cal
L}[\phi^{-},\psi^{-}] +J_{+}^{\phi}\phi^{+} +
J_{-}^{\phi}\phi^{-} + [J_{+}^{\psi} \psi^{+}] +
[J_{-}^{\psi}\psi^{-}] \right)\right]  %\nonumbe
\end{eqnarray}
where we have abbreviated $[{\cal D}\psi]={\cal D}\bar{\psi}{\cal D}\psi$, $[J_{+}^{\psi} \psi^{+}]= J_{+}^{\psi} \psi^{+}+\bar{\psi}^{+}\bar{J_+}^{\bar\psi^+} $ for the sake of notational convenience. Note that the Schwinger-Keldysh formalism requires two kind of fields. The $+$ type  is responsible for forward time evolution whereas the $-$ type  for the backward evolution. The latter kind of fields represent virtual particles only. Each type of field is associated with a respective  source, $J$. Also, the $\delta$-function appearing in \ref{Z:inin} guarantees that the field configurations for the $\pm$ types remain the same on the final hypersurface at $\eta=\eta_f$.
 
  The various $n$-point correlation functions are obtained by functionally differentiating \ref{Z:inin} with respect to the source terms. For example for the scalar field we have 
\begin{eqnarray} \label{npointfunctions}
\left. \mathrm{Tr}\left[ {\rho}(\eta_{0})
\overline{T}[\phi(x_1)\dots \phi(x_n)]
T[\phi(y_1)\dots \phi(y_k)] \right] =
\frac{\delta^{n+k}{\cal Z}[J, \rho(\eta_0)] } {i\delta \!
J_{-}^{\phi}(x_{1})\cdots i\delta\! J_{-}^{\phi}(x_{n})
i\delta\! J_{+}^{\phi}(y_{1})\cdots i\delta\!
J_{+}^{\phi}(y_{k})} \right|_{J = 0}  
\end{eqnarray}

Using the above equation, we obtain  the following four scalar propagators
\begin{subequations}
\label{propagators}
\begin{eqnarray}
i\Delta^{++}_{\phi}(x,x^\prime) &=&
\mathrm{Tr}\left[{\rho}(\eta_{0})
T[\phi(x^\prime)\phi(x)] \right] =
 \mathrm{Tr}\left[{\rho}(\eta_{0}) \phi^+(x)\phi^+(x^\prime)\right] =
\left.\frac{\delta^2{\cal Z}[J, \rho(\eta_0)]} {i\delta\!
J_{+}^{\phi}(x) i\delta\! J_{+}^{\phi}(x^\prime)}
\right|_{J=0}
\label{propagatorsa} \\
i\Delta^{--}_{\phi}(x,x^\prime) &=&
 \mathrm{Tr}\left[{\rho}(\eta_{0}) \overline{T} [ \phi(x^\prime)\phi(x)]
\right] =
 \mathrm{Tr}\left[{\rho}(\eta_{0}) \phi^-(x^\prime)\phi^-(x)\right] =
\left. \frac{\delta^2{\cal Z} [J, \rho(\eta_0)]} {i\delta\!
J_{-}^{\phi}(x)i\delta\! J_{-}^{\phi}(x^\prime)}
\right|_{J=0}
\label{propagatorsb} \\
i\Delta^{-+}_{\phi}(x,x^\prime) &=&
 \mathrm{Tr}\left[{\rho}(\eta_{0})  \phi(x)\phi(x^\prime)\right] =
 \mathrm{Tr}\left[{\rho}(\eta_{0}) \phi^-(x)\phi^+(x^\prime)\right] =
\left.\frac{\delta^2{\cal Z} [J, \rho(\eta_0)] } {i\delta\!
J_{-}^{\phi}(x)i\delta\! J_{+}^{\phi}(x^\prime)}
\right|_{J=0} \label{propagatorsc}
\\
i\Delta^{+-}_{\phi}(x,x^\prime) &=&
 \mathrm{Tr}\left[{\rho}(\eta_{0})\phi(x^\prime)\phi(x)\right] =
 \mathrm{Tr}\left[{\rho}(\eta_{0})\phi^-(x^\prime)\phi^+(x)\right]=
\left.\frac{\delta^2{\cal Z} [J, \rho(\eta_0)] } {i\delta\!
J_{+}^{\phi}(x)i\delta\! J_{-}^{\phi}(x^\prime)}
\right|_{J=0}
\label{propagatorsd}
\end{eqnarray}
\end{subequations}
where we have taken $\eta\gtrsim \eta'$ above.
These propagators can be identified  as the Feynman and anti-Feynman propagators for the time ordered and anti-time ordered cases, respectively, and the rest are the two Wightman functions. We  write
\begin{eqnarray}
\label{propagatoridentities}
i\Delta^{++}_{\phi}(x,x^\prime) &=& \theta(\eta-\eta^\prime)i
\Delta^{-+}_{\phi}(x,x^\prime) + \theta(\eta^\prime-\eta)i
\Delta^{+-}_{\phi}(x,x^\prime) \nonumber
\\
i\Delta^{--}_{\phi}(x,x^\prime) &=& \theta(\eta^\prime-\eta)i
\Delta^{-+}_{\phi}(x,x^\prime) + \theta(\eta-\eta^\prime)i
\Delta^{+-}_{\phi}(x,x^\prime) 
\end{eqnarray}
It is also easy to see that 
\begin{eqnarray}
i\Delta^{++}_{\phi}(x,x^\prime) +
i\Delta^{--}_{\phi}(x,x^\prime) &=& i
\Delta^{-+}_{\phi}(x,x^\prime) +i
\Delta^{+-}_{\phi}(x,x^\prime) \label{propagatoridentitiesc}
\\
i \Delta^{-+}_{\phi}(x,x^\prime)&=&i
\Delta^{+-}_{\phi}(x^\prime,x) \label{propagatoridentitiesd}
\end{eqnarray}
For our purpose, we also define  the spectral and the statistical  propagators respectively as,
\begin{eqnarray} \label{defintion of scalar correlators:2}
&&i\Delta^{c}_{\phi} (x,x^\prime) = \mathrm{Tr}\left(
{\rho}(\eta_{0})  [\phi(x),\phi(x^\prime)]\right)= i
\Delta^{-+}_{\phi}(x,x^\prime) - i
\Delta^{+-}_{\phi}(x,x^\prime)\nonumber\\
\label{Hadamard}
&&F_{\phi}(x,x') = \frac{1}{2} \mathrm{Tr}\left[ {\rho}(\eta_{0})
[ \phi(x),\phi(x')]_+ \right]=
\frac{1}{2}\Big(i\Delta^{-+}_{\phi}(x,x') +
i\Delta^{+-}_{\phi}(x,x')\Big) \label{statistical}
\end{eqnarray}
Owing to the spacelike translational invariance of the de Sitter spacetime, \ref{l0}, it is often convenient to introduce a $3$-momentum space, via the Fourier transform  
\begin{eqnarray}
f(\eta, \eta^{\prime},k) &=& \int d^3\vec{r} \,{\rm  e}^{- i \vec k\cdot \vec{r} } f(x,x^\prime) 
\end{eqnarray}
where $k=|\vec{k}|$ and $\vec{r}= \vec{x}-\vec{x}'$. In terms of this, it is easy to write from \ref{defintion of scalar correlators:2}
\begin{eqnarray}\label{defintion of scalar correlators}
i \Delta_{\phi}^{\mp \pm}(\eta, \eta^{\prime},k) &=& F_{\phi}(\eta,\eta^\prime,k) \pm \frac{i}{2}\Delta_{\phi}^c(\eta,\eta^\prime,k)\nonumber\\
i\Delta^{\pm \pm}_{\phi}(\eta, \eta^{\prime},k) &=& F_{\phi}(\eta, \eta^{\prime},k) 
      \pm \frac{1}{2} \text{sign} (\eta - \eta') i\Delta^c_{\phi}(\eta, \eta^{\prime},k)
\end{eqnarray}    
which will be useful for our future purpose. 

%%%%%%%%%%%%
\section{Some useful identities \label{defAndConv}}
%%%%%%%

We have to frequently use the following special functions for our present purpose \cite{Gr, Ar},
\begin{eqnarray}\label{ci,si}
\text{si} (z)&=&-\int_{z}^{\infty}\frac{\sin t \,dt}{t}=
\int_{0}^{z}\frac{\sin t \,dt}{t}-\frac{\pi}{2}
%=\text{Si}(z)-\frac{\pi}{2}
\, 
\label{sine integral}\\
\text{ci}(z)&=&-\int_{z}^{\infty}\frac{\cos t\,dt}{t}
\, 
\label{cos integral}
%\text{E}_1 ( i x)&=& - \gamma_E - \ln (i x) - \sum_{k=1}^{\infty} \frac{(-ix)^k}{k k!}  =  i  \text{si} ( x)-\text{ci} ( x)\,, \quad x>0\, \\
%\text{Ein} (z) &=& \int_0^z \frac{1-e^{-t}}{t} dt = \sum_{n=1}^{\infty} \frac{(-1)^{n-1}z^n }{n! n}  =  E_1 (z) + \ln z + \gamma_E
\end{eqnarray}
where  the first and the second  are respectively the sine and cosine  integral functions. 

We shall use some  relations derived in~\cite{Prokopec:2008gw} for the complexified de Sitter invariant $2$-point functions, $y_{\pm\pm}$, defined in \ref{distance}, \ref{distance1} and \ref{distance2}. For example, we have 
\begin{eqnarray}
 \Big( {\frac{y_{\pm \pm}}{4}}\Big)^{1-{d}} = \Big[\frac{2}{(d-2)^2} \frac{\square}{H^2} -\frac{2}{(d-2)}\Big] \Big( {\frac{y_{\pm \pm}}{4}}\Big)^{2-{d}} 
\label{1}   
\end{eqnarray}
We also have 
\begin{multline}
\Big( {\frac{y_{\pm \pm}}{4}}\Big)^{2-{d}} = \Big[\frac{2}{(d-3)(d-4)} \frac{\square}{H^2} -\frac{d(d-2)}{2 (d-3)(d-4)}   + \frac{d-6}{2(d-3)} \Big]\Big( {\frac{y_{\pm \pm}}{4}}\Big)^{3-{d}} \\
-\Big[ \frac{2}{(d-3)(d-4)} \frac{\square}{H^2} - \frac{d(d-2)}{2 (d-3)(d-4)} \Big]\Big( {\frac{y_{\pm \pm}}{4}}\Big)^{1-d/2} \pm \frac{2(4 \pi)^{d/2}}{(d-3)(d-4)\Gamma\big[\frac{d}{2}-1 \big]} \frac{i \delta^d (x-x^{\prime})}{(Ha)^d}\label{deSitterRel}
\end{multline}
as well as
\begin{eqnarray}
 \frac{\square}{H^2}\Big( {\frac{y_{\pm \pm}}{4}}\Big)^{1-d/2} = \pm \frac{(4 \pi)^{d/2}}{\Gamma\big[\frac{d}{2}-1 \big]} \frac{i \delta^d (x-x^{\prime})}{(Ha)^d} + \frac{d(d-2)}{4}\Big( {\frac{y_{\pm \pm}}{4}}\Big)^{1-d/2}
\end{eqnarray}
Introducing a renormalisation scale $\mu$ with mass dimension one, we can modify \ref{deSitterRel} by including a term dependent on $\mu$. This modification allows us to eliminate the divergence in the self energy using appropriate counter terms in the action.
% given by $(d-4)^{-1}\mu^{d-4} a^{-d} \delta^d \big(x-x^{\prime} \big)$. 
Additionally, we have
\begin{eqnarray}
  \Big( {\frac{y_{\pm \pm}}{4}}\Big)^{3-{d}} = \Big( {\frac{y_{\pm \pm}}{4}}\Big)^{1-d/2}\Big[1- \frac{d-4}{2}\ln   y_{\pm \pm}  + \mathcal{O}(d-4)^2  \Big]  
\end{eqnarray}
Putting things together now, we have from \ref{deSitterRel},
\begin{multline}
\Big( {\frac{y_{\pm \pm}}{4}}\Big)^{2-{d}} =
 \pm \frac{2 (4 \pi )^{d/2} }{(d-3)(d-4)\Gamma\big[\frac{d}{2}-1\big] } \Big(  \frac{ \mu}{H}\Big) ^{d-4}  \frac{i \delta^d \big(x-x^{\prime} \big)}{(Ha)^d}
-  \frac{\square}{H^2} \Big(\frac{4}{y_{\pm \pm}}\ln \frac{\mu^2 y_{\pm \pm}}{H^2} \Big) + \frac{4}{y_{\pm \pm}} \Big(2 \ln  \frac{\mu^2 y_{\pm \pm} }{H^2} -1 \Big)  + \mathcal{O}(d-4)
\end{multline}

We also note that for any function $f(y)$, not containing $y^{-1}$, we have 
\begin{align}\label{i1}
\frac{\square}{H^2} f(y) = (4-y) y f^{\prime \prime} (y) + 4(2-y)f^{\prime}(y)
\end{align}
where a prime denotes derivative once with respect to $\eta$. The above gives identities useful for our purpose
\begin{eqnarray}\label{i2}
\frac{1}{y} &=& \frac{1}{4} \frac{\square}{H^2} \ln y + \frac{3}{4}\, \\
\frac{\ln y}{y}   &=&
  \frac{1}{8} \frac{\square}{H^2} \Big[ \ln^2 y - 2  \ln y  \Big]
  + \frac{3}{4} \ln  y  	
- \frac{1}{2}\,
\end{eqnarray}

We also note  from \ref{distance1}, \ref{distance2} that as $\e\to 0$
\begin{eqnarray}
&&\ln \big( \Delta x_{\mp \pm}^2 \big) = \ln  \big|\Delta \eta^2 - r^2 \big| 
 \pm  i \pi \, \text{sign}\big(\eta,\eta^{\prime} \big) \theta \big( \Delta \eta^2 - r^2 \big) \nonumber\\
&& \ln \big( \Delta x_{\pm \pm}^2 \big) = \ln  \big|\Delta \eta^2 -r^2 \big| 
 \pm  i \pi \,  \theta \big( \Delta \eta^2 - r^2 \big) 
\end{eqnarray}\label{log}
where $\text{sign}(\eta,\eta^{\prime})$ is the usual signum function.

For our purpose of solving the Kadanoff-Baym equations, \ref{2PF}, it will be convenient  to define \cite{Friedrich:2019hev},
\begin{eqnarray}\label{newenergies}
M^{F}(\eta, \eta^{\prime}) &:=& \frac{1}{2}\Big[M^{++}(\eta, \eta^{\prime})+M^{--}(\eta, \eta^{\prime}) \Big] = \text{Re} \, M^{++}(\eta, \eta^{\prime})\, , \\
M^{c}(\eta, \eta^{\prime}) &:=&-i\,\text{sign} \big(\Delta\eta \big) \Big[M^{++}(\eta, \eta^{\prime})-M^{--}(\eta, \eta^{\prime})\Big] =2 \, \text{sign} \big(\Delta\eta \big)\,  \text{Im} \, M^{++}(\eta, \eta^{\prime})\label{newenergies1}
\end{eqnarray}
where $aa'M^{++}$ and $aa'M^{--}$ are respectively the self energies corresponding to the Feynman and anti-Feynman propagators. We also note
\begin{eqnarray}\label{energies}
 \Big[M^{++}(\eta, \eta^{\prime})-M^{--}(\eta, \eta^{\prime}) \pm \Big(M^{-+}(\eta, \eta^{\prime})-M^{+-}(\eta, \eta^{\prime})\Big)\Big]  &= &\pm 2 \theta(\pm \Delta \eta )
 \, i M^{c}  (\eta, \eta^{\prime}) \\
 \Big[M^{++}(\eta, \eta^{\prime})+M^{--}(\eta, \eta^{\prime})\Big] +
  \text{sign} (\eta  \!-\!  \eta^{\prime})  \Big[M^{-+}(\eta, \eta^{\prime})+M^{+-}(\eta, \eta^{\prime})\Big]&= & 4 \theta(\eta \!-\! \eta^{\prime})\, M^{F}(\eta, \eta^{\prime})\label{energies1}
\end{eqnarray}
where $aa'M^{-+}$ or $aa'M^{+-}$ are the self energies corresponding to the Wightman functions. 

%%%%%%
\section{Fourier transform of logarithms \label{fourier}}
%%%%%%%%

In order to take \ref{inhomSelfMass1} to momentum space, \ref{M++Full}, we show in this appendix
\begin{multline}
\int d^{3}\vec{r}\, e^{-i \vec{k} \cdot \vec{r}}\Bigg[\frac{1}{2}\ln^2 \frac{y_{++}}{4} + f\big(\eta, \eta^{\prime} \big)\ln \frac{y_{++}}{4}  \Bigg] 
\\ =
-\frac{4 \pi^2}{k^3} \Bigg[2+ \big[1+ i k |\Delta \eta | \big] \Big(  \ln \frac{a a^{\prime} H^2| \Delta \eta|}{2k} +  \frac{ i\pi}{2}- \gamma_E + f(\eta, \eta^{\prime})\Big) \Bigg] e^{-i k |\Delta \eta|} \\
+\frac{4 \pi^2}{k^3} \big(1 - i k |\Delta \eta| \big)\Bigg[ \text{ci} \big[ 2 k| \Delta \eta|  \big]  -i \,  \text{si} \big[ 2 k |\Delta \eta|  \big]  \Bigg] e^{+i k |\Delta \eta|}\label{fTlogsApp}
\end{multline}
where $\Delta \eta  = \eta - \eta^{\prime}$ and $f(\eta, \eta^{\prime})$ is some $k$-independent function. Above equation was first derived in \cite{Friedrich:2019hev}, we briefly review it here for the sake of completeness. In order to solve above equation, we require integrals of the following type,
\begin{align}
 \mathcal{I}_n (x)  & \equiv x^2 \int_0^{\infty} dz \, z \sin \big[x z  \big] { \ln^n \Big( |1- z^2  |\Big) }
 \\
 &= x^2 \Bigg[ \frac{d^n}{db^n} \int_0^{\infty} dz \, z \sin \big[x z  \big]  |1- z^2  |^{b}  \Bigg]_{{b}=0} 
 \end{align}
 Using 
 \begin{multline}
  \int_0^{\infty} dz \, z \sin (x z)  |1- z^2  |^b =  \frac{\sqrt{\pi}}{2} \Big( \frac{2}{x} \Big)^{b+ \frac{1}{2}} \Gamma \big[ b+1\big] \Big[ J_{b + \frac{3}{2}} \big(x \big) + Y_{-b - \frac{3}{2}} \big(x \big) \Big]\, \,\,\, (x> 0, \quad  -1< b < 0)
 \end{multline}
 where $J_{n}\, , Y_{m}$ are the Bessel functions of the first and second kind respectively, we find 
 \begin{align}
\mathcal{I}_{1} \big(x \big)   =  - \pi \Big[\cos x + x \sin x   \Big]
 \end{align}
 and 
  \begin{multline}
\mathcal{I}_{2}(x )   =   2 \pi \left[-2\cos x +   ( \cos x + x \sin x ) \left(  \text{ci}(2 x)  + \gamma_E - \ln \frac{2}{x} \right) +\left(    \sin x   -x \cos x \right)  \text{si}(2 x)\right]%\nonumber\\
 \end{multline}
 where the sine and cosine integral functions are defined in \ref{sine integral} and \ref{cos integral} in the preceding appendix and $\gamma_E$ is the Euler constant.
 
Using \ref{log}, we establish
\begin{multline}
\label{log1}
\int d^{3}\vec{r} e^{-i \vec{k} \cdot\vec{r}} \ln  \frac{y_{++}}{4}   =  \frac{4 \pi }{k} \int_0^{\infty} dr \, r \sin k r \ln   \frac{y_{++}}{4}  \\
=   \frac{4 \pi }{k} \int_0^{\infty} dr \, r \sin k r \ln  |1- {r^2}{\Delta \eta^{-2} }|  
+   \frac{4 \pi^2 i}{k} \int_0^{\infty} dr \, r \sin k r\, \theta \big( \Delta \eta^2 - r^2  \big)  
\\
=   \frac{4 \pi }{k^3} ( k\Delta \eta  )^2 \int_0^{\infty} dz \, z \sin ( k | \Delta \eta | z  ) \ln  |z^2 - 1  |  
+   \frac{4 \pi^2 i}{k} \int_0^{| \Delta \eta | } dr \, r \sin k r   
 = -\frac{4 \pi^2}{k^3} \big[1+i k |\Delta \eta|  \big]e^{- i k |\Delta \eta| } 
\end{multline}
We also compute
\begin{multline}
\int d^{3}\vec{r} e^{-i \vec{k} \cdot\vec{r}} \ln^2 \frac{y_{++}}{4} 
=  \frac{4 \pi }{k} \int_0^{\infty} dr \, r \sin k r\, \ln^2  |1- {r^2}{\Delta \eta^{-2} }  | 
 + \frac{8 \pi }{k} \ln  \frac{a a^{\prime} H^2 \Delta \eta^2}{4}  \int_0^{\infty} dr \, r \sin k r\, \ln  |1- {r^2}{\Delta \eta^{-2} } | \\ 
+   \frac{8 \pi^2 i}{k} \int_0^{\infty} dr \, r \sin k r\, \ln  |1- {r^2}{\Delta \eta^{-2} }   |\theta \big( \Delta \eta^2 - r^2  \big) 
-\frac{4 \pi^3 }{k}\int_0^{\infty} dr \, r \sin k r\,\theta \big( \Delta \eta^2 - r^2  \big) \\
+ \frac{8 \pi^2 i}{k}\ln\frac{a a^{\prime} H^2 \Delta \eta^2}{4} \int_0^{\infty} dr \, r \sin k r\, \theta \big( \Delta \eta^2 - r^2  \big)   \\
=  \frac{4 \pi }{k^3}( k\Delta \eta )^2 \int_0^{\infty} dz \, z \sin k |\Delta \eta | { \ln^2  |1- z^2  | }
 + \frac{8 \pi }{k^3} \ln \frac{a a^{\prime} H^2 \Delta \eta^2}{4}  (k\Delta \eta)^2 \int_0^{\infty} dz \, z \,\sin (k |\Delta \eta |z ) \ln |1- z^2  | \\ 
+  \frac{8 \pi^2i }{k^3} ( k\Delta \eta)^2 \int_0^{1} dz \, z\, \sin (k |\Delta \eta | z ) \ln  |1- z^2  |
-\frac{4 \pi^2 }{k} \Bigg[\pi - 2i\ln  \frac{a a^{\prime} H^2 \Delta \eta^2}{4}  \Bigg]\int_0^{|\Delta \eta |} dr \, r \sin k r   \\
= -\frac{8 \pi^2}{k^3} \Bigg[2+ \big[1+ i k |\Delta \eta | \big] \Big(  \ln \frac{a a^{\prime} H^2| \Delta \eta|}{2k} +  \frac{i \pi}{2}- \gamma_E \Big) \Bigg] e^{-i k |\Delta \eta|} 
+\frac{8 \pi^2}{k^3} \big(1 - i k |\Delta \eta| \big)\Bigg[ \text{ci} \big[ 2 k| \Delta \eta|  \big]  -i \,  \text{si} \big[ 2 k |\Delta \eta|  \big]  \Bigg] e^{+i k |\Delta \eta|}\label{log2}
\end{multline}
We next combine the results \ref{log1} and \ref{log2} in order to finally obtain \ref{fTlogsApp}. One can similarly find the results  for $y_{--}$ and $y_{+-}$ and $y_{-+}$.

%%%%%%%%%%%
\section{Tree level infrared correlators }\label{corr1}
%%%%%%%%%%%%
The infrared (IR) effective field theory in de Sitter pertains to the super-Hubble modes that have been redshifted over time, and is free from any ultraviolet divergences. This field theory is composed of truncated IR modes, expressed as
\begin{eqnarray}
\phi(\eta, \vec{x})  = \int \frac{d^3 \vec{k}}{(2\pi)^{3/2}} \theta (Ha-k) \left[a_{\vec k}\, u(k,\eta) e^{-i\vec{k}\cdot \vec{x}}+a^{\dagger}_{\vec k}\, u^{\star}(k,\eta) e^{i\vec{k}\cdot \vec{x}}\right]
\label{c1}
\end{eqnarray}
where $u(\vec{k},\eta)$, the Bunch-Davies mode function, is given by \ref{modfunc}.
By taking the limit $\eta \to 0^-$, the super-Hubble IR modes (where $k$ is restricted by a cut off) can be expanded as 
\begin{eqnarray}
u(\vec{k},\eta) \vert_{\rm IR} \approx \frac{H}{\sqrt{2} k^{3/2}}\left[1+\frac12 \left(\frac{k}{Ha} \right)^2 + \frac{i}{3}\left(\frac{k}{Ha} \right)^3 +\,{\rm subleading~terms } \right]
\label{c2}
\end{eqnarray}
Thus the leading temporal part of the Bunch-Davies modes becomes nearly a constant in this limit. Substituting this into \ref{c1} and dropping  the suffix `IR' without any loss of generality, we have
\begin{eqnarray}
\phi(\eta, \vec{x})  =\frac{H}{\sqrt{2}} \int \frac{d^3 \vec{k}}{(2\pi)^{3/2}} \frac{\theta (Ha-k)}{k^{3/2}} \left[a_{\vec k}\,  e^{-i\vec{k}\cdot \vec{x}}+a^{\dagger}_{\vec k}\,  e^{i\vec{k}\cdot \vec{x}}\right]+\,{\rm subleading~terms}
\label{c3}
\end{eqnarray}
The step function appearing above ensures that we are essentially dealing with long wavelength modes. 

The IR Wightman functions in terms of these modes are accordingly given by (cf., \ref{Propagators in the Schwinger-Keldysh Formalism})
\begin{eqnarray}
i \Delta^{-+}_{\phi}(x,x^\prime)&&=\frac{H^2}{2} \int \frac{d^3 {\vec k}}{(2\pi)^3 k^3} e^{i{\vec k}\cdot({\vec x}-{\vec y})}\theta(Ha-k)\theta(Ha'-k)(1+ik\eta)(1-ik\eta')e^{-ik(\eta -\eta')}\nonumber\\ &&= \int \frac{d^3 {\vec k}}{(2\pi)^3} e^{i{\vec k}\cdot({\vec x}-{\vec y})}i \Delta_{-+}(k,\eta,\eta')\nonumber\\
i \Delta^{+-}_{\phi}(x,x^\prime) &&=\frac{H^2}{2} \int \frac{d^3 \vec{k}}{(2\pi)^3 k^3} e^{i\vec{k}\cdot(\vec{x}-\vec{y})}\theta(Ha-k)\theta(Ha'-k)(1-ik\eta)(1+ik\eta')e^{ik(\eta -\eta')}\nonumber\\&&= \int \frac{d^3 {\vec k}}{(2\pi)^3} e^{i{\vec k}\cdot({\vec x}-{\vec y})}i \Delta_{+-}(k,\eta,\eta') 
\label{nc2}
\end{eqnarray}
Using now \ref{c2}, we  compute
\begin{eqnarray}
 i \Delta^{+-} _\phi (\eta',\eta'', k)\vert_{ k | \Delta \eta| \ll 1}&&\approx  \frac{H^2 \theta(Ha'-k)\theta(Ha''-k)}{2k^3} \left(1+\frac{ik^3}{3H^3 a''^3} \right)\qquad (\eta' \gtrsim \eta'')
\label{nc4'}
\end{eqnarray}
and  $i \Delta^{+-}=(i \Delta^{-+})^{\star}$.  We also have for our purpose from \ref{statistical}
\begin{eqnarray}
\Delta_{\phi}^c (\eta, \eta^{\prime}, k )_{ k | \Delta \eta| \ll 1}&& = i \Delta^{-+} _\phi (\eta,\eta',k) - i \Delta^{+-} _\phi (\eta,\eta',k) \approx  -\frac{i \theta(Ha-k)\theta(Ha'-k)}{3H a'^3}\nonumber\\
 F_{\phi} (\eta, \eta^{\prime}, k )_{ k | \Delta \eta| \ll 1}  &&= \frac12\left(i \Delta^{+-} _\phi (\eta,\eta', k) + i \Delta^{-+} _\phi (\eta,\eta', k)\right)\approx \frac{H^2}{2k^3}\theta(Ha-k)\theta(Ha'-k)
\label{nc4}
\end{eqnarray}
%
%%%%%%%%%%%%%%%%%%%%

\end{document}